\documentstyle[aas2pp4]{article} 

\def\comment#1{{}}

\begin{document}

\title{Discovery of a Hard X-Ray Source, SAX~J0635+0533, in the
Error Box of the Gamma-Ray Source 2EG~0635+0521}

\author{Philip Kaaret\altaffilmark{1}, Santina
Piraino\altaffilmark{2}, Jules Halpern\altaffilmark{3,4}, and
Michael Eracleous\altaffilmark{5,4}}

\authoremail{pkaaret@cfa.harvard.edu}

\altaffiltext{1}{Harvard-Smithsonian Center for Astrophysics,
60 Garden St., Cambridge, MA 02138} 

\altaffiltext{2}{Istituto di Fisica Cosmica ed Applicazioni
dell'Informatica, CNR, Via Ugo La Malfa 153, 90146 Palermo,
Italy and Istituto di Fisica, Universita' di Palermo, via
Archirafi 36, 90123  Palermo,  Italy } 

\altaffiltext{3}{Columbia Astrophysics Laboratory, Columbia
University, New York, NY 10027}

\altaffiltext{4}{Visiting Astronomer, Kitt Peak National
Observatory, National Optical Astronomy Observatories, which
is operated by AURA, Inc., under a cooperative agreement with
the National Science Foundation.}

\altaffiltext{5}{Hubble Fellow, Department of Astronomy,
University of California, Berkeley, CA 94720-3411. Current
address: Department of Astronomy and  Astrophysics, The
Pennsylvania State University, 525 Davey Lab, University Park,
PA 16802.}

\begin{abstract}

We have discovered an x-ray source, SAX~J0635+0533, with a hard spectrum
within the error box of the GeV gamma-ray source in Monoceros,
2EG~J0635+0521.  The unabsorbed flux from the source is $1.2 \times 10^{-11}
\rm \, erg \, cm^{-2} \, s^{-1}$ in the 2-10~keV band.  The x-ray spectrum is
consistent with a simple powerlaw model with absorption.  The photon index
is $1.50 \pm 0.08$ and we detect emission out to 40~keV.  Optical
observations identify a counterpart with a V-magnitude of 12.8.  The
counterpart has broad emission lines and the colors of an early B type star.
If the identification of the x-ray/optical source with the gamma-ray source
is correct, then the source would be a gamma-ray emitting x-ray binary.

\end{abstract}

\keywords{pulsars: general --- gamma rays: observations ---
stars: individual (2EG J0635+0521, SAX~J0635+0533,
LSI~61$^{\circ}$~303, 2CG~135+01, PSR~J1259-63) --- stars:
neutron}

\section{Introduction}

The unidentified gamma-ray point sources concentrated along
the Galactic plane present a long outstanding puzzle
(Swanenburg et al. 1981).  Their positions, timing, and energy
spectra should provide a means for their identification. 
However, the gamma-ray error boxes are sufficiently large that
multiple optical and x-ray candidates are usually found for
Galactic sources.  The only convincing identifications have
been made through detection of a pulsed gamma-ray signal at a
period known from radio or x-ray observations (Bignami \&
Hermsen 1983, Halpern \& Holt 1992, Thompson et al. 1994,
Ramanamurthy et al.  1995).  The known gamma-ray pulsars have
spectra significantly harder than those of most of the
unidentified sources.  Recently, Merck et al. (1996) studied
the energy spectra of the unidentified Galactic gamma-ray
sources and identified only eight sources with energy spectra
sufficiently hard to be pulsar candidates.

Here, we report on x-ray and optical observations in the field
of one of the unidentified Galactic gamma-ray sources with a
hard gamma-ray spectrum, 2EG~J0635+0521.  In \S 2, we report
the discovery of an x-ray point source with an unusually hard
spectrum within the error box of 2EG~J0635+0521.  In \S \S 3
and 4, we describe the properties of an optical counterpart of
the x-ray source.  We conclude, in \S 5, with comments on
identification of the x-ray and gamma-ray sources and on the
possible nature of the source.

\begin{figure}[tb] \epsscale{0.9} \plotone{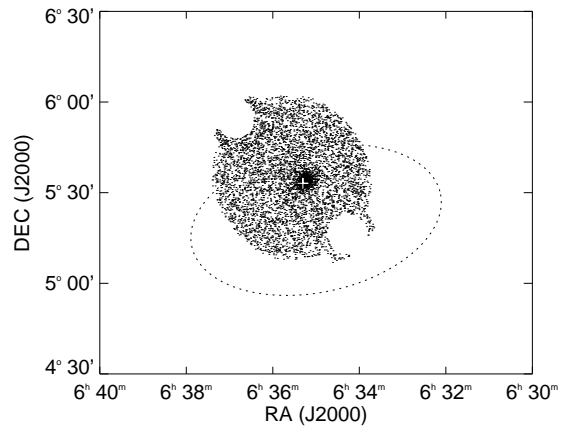}
\figcaption{X-ray image of SAX~J0635+0533 for the 5--12~keV
band.  Shown are the position of the optical counterpart
(cross), and the 95\% confidence error ellipse for
2EJ~0635+0521 (dashed line).} \label{fig_image} \end{figure}

\begin{figure}[tb] \epsscale{0.7} \plotone{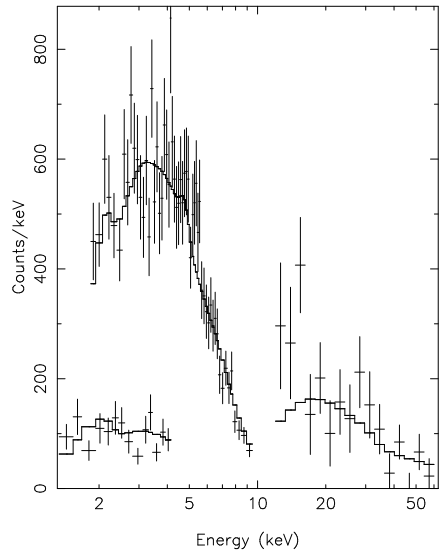}
\figcaption{X-ray spectrum of SAX~J0635+0533.  Shown are data
from the LECS (1.0-4.0~keV), MECS (1.8--10~keV), and PDS
(12--60~keV).  The solid line is a fit of a powerlaw of photon
index of 1.5 with absorption.} \label{fig_xray_spectrum}
\end{figure}

\section{X-ray observations}

Using the {\it Satellite Italiano per Astronomia X}
(BeppoSAX), we observed a field centered on a region of
unusually high x-ray hardness and brightness found in an x-ray
survey of the Monoceros supernova remnant (Leahy, Naranan, \&
Singh 1986) and located within the error box of the EGRET
source 2EG~J0635+0521 (Thompson et al. 1995).  The observation
was carried out from 23 Oct 1997 3:20:26 UT to 24 Oct 1997
0:11:52 UT.  The field of view of the BeppoSAX narrow field
instruments covered approximately half of the error box of
2EG~J0635+0521.  The x-ray images from the LECS and the MECS 2
and 3 detectors reveal a single point source located at RA =
$\rm 06^{h} 35^{m} 17^{s}.4$, DEC = $+05^{\circ} 33' 20''.9$
(J2000).  There is a 1' systematic uncertainty in the position
reconstruction.  We denote the source as SAX~J0635+0533.  This
source lies within the error box of 2EG~J0635+0521 (see
Figure~1) and near the peak of the gamma-ray emission as
reconstructed by Jaffe et al. (1997).  The image is consistent
with that of a single point source, see Fig.~1. We find no
evidence for diffuse hard x-ray emission.

The x-ray spectrum of the source is quite hard with a spectral
index near 1.5 and significant emission detected out to
40~keV.  Shown in Figure~2 are data from the LECS, the
combined MECS 2 and 3, and the PDS.   We used an extraction
radius of 2' for the MECS and 4' for the LECS, the standard
background files, and response matrixes calculated for the
source position 3' off-axis.  The x-ray spectrum from all the
instruments can be fitted with a simple powerlaw model with
absorption.  This fit is acceptable with $\chi^2/{\rm DoF} =
97/81$.  The fitted column density is $(2.0 \pm 0.3) \times
10^{22} \rm \, cm^{-2}$ and the powerlaw index is $1.50 \pm
0.08$.  The unabsorbed flux from the source is $1.2 \times
10^{-11} \rm \, erg \, cm^{-2} \, s^{-1}$ in the 2-10~keV
band.  We also fitted an exponentially cutoff powerlaw which
gave $\chi^2/{\rm DoF} = 96/80$.  The exponential cutoff does
not improve the fit.  The 90\% confidence lower bound on the
cutoff energy is 37~keV.  We also fitted a Raymond-Smith model
spectrum which gave a worse fit $\chi^2/{\rm DoF} = 122/83$
and an unphysically high temperature, $kT > 50 \rm \, keV$.

We searched for pulsations in the x-ray data over a period
range from 0.030~s to 1000~s.  We found no significant pulsed
signal. From the period with the highest single trial
significance, we place an upper limit on the pulsed fraction
of 26\%.  The continuum of the timing power spectrum is
consistent with the Poisson level for frequencies above
0.002~Hz.

This field was previously viewed with the Einstein IPC (region
3 in Leahy et al. 1986).  The source was not detected with
Einstein because, we now know, it fell on a strongback arm. 
This field was also observed for 4371~s by the {\it ROSAT\/}
PSPC on 1992 October~9.  The position of SAX~J0635+0533 falls
$34^{\prime}$ off axis in this image, where the point-spread
function is significantly degraded.  Nevertheless, there may
be a very weak source present at this position.  Folding the
spectrum of SAX~J0635+0533 through the {\it ROSAT\/} response,
we find that the predicted number of counts is roughly
consistent with the number observed.  There is no strong
evidence for variability of the source.

\begin{figure}[tb] \epsscale{0.7} \plotone{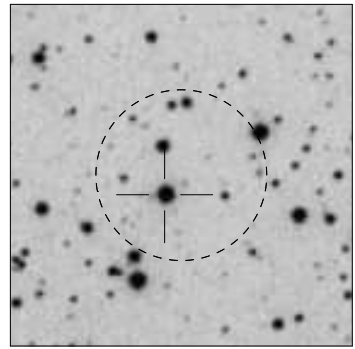}
\figcaption{Finding chart for SAX~J0635+0533 from the
Digitized Sky Survey.  Field is $4^{\prime}\times 4^{\prime}$.
North is up, east is left. The X-ray error circle has a radius
of $1^{\prime}$. The position of the optical counterpart is
(J2000) RA = $06^{\rm h}35^{\rm m}18^{\rm s}\!.29$, Dec =
$+05^{\circ}33^{\prime}6^{\prime\prime}\!.3$.}
\label{fig_finding_chart} \end{figure}

\begin{figure*}[tb] \epsscale{1.3} \plotone{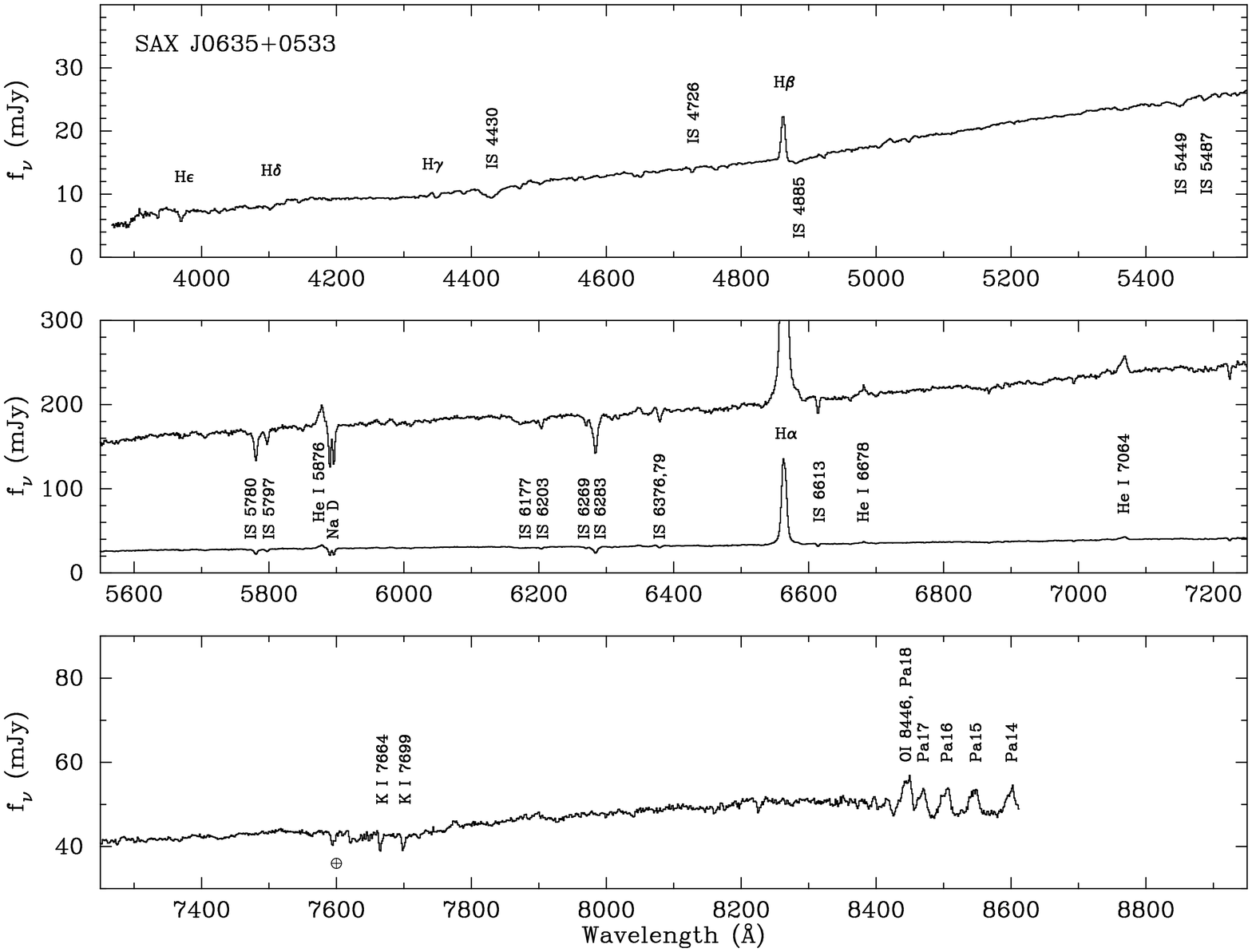}
\figcaption{Composite optical spectrum of SAX~J0635+0533
obtained on the KPNO 2.1m telescope on 1998 January 29 and
30.  In the middle panel, the flux scale refers to the lower
trace.  The upper trace is the same spectrum multiplied by a
factor of 6.} \label{fig_optical_spectrum} \end{figure*}

\begin{figure}[tb] \epsscale{0.7} \plotone{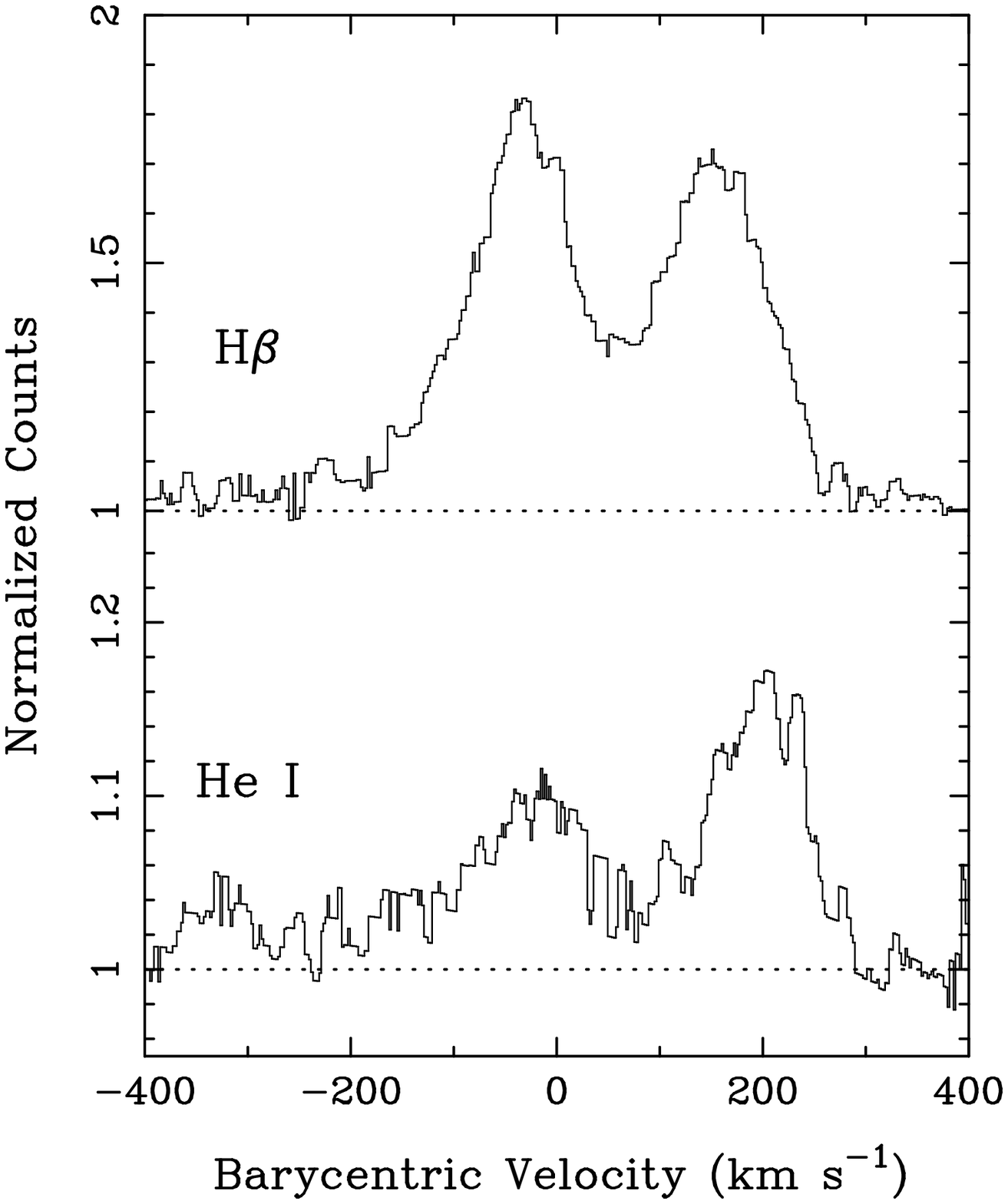}
\figcaption{Keck HIRES spectrum of selected emission lines,
H$\beta$ and He~I~$\lambda$5876, normalized to the continuum
and smoothed with a 14~km~s$^{-1}$ running boxcar filter.}
\label{fig_keck_spectrum} \end{figure}

\section{Optical observations}

In order to identify the optical counterpart of
SAX~J0635+0533, we obtained optical spectra of seven stars
within and around the $1^{\prime}$ radius error circle on 1997
Dec. 10, using the 2.4~m Hiltner telescope of the MDM 
Observatory.  The spectra covered the wavelength range
4500--7500~\AA\ at 4~\AA\ resolution. The brightest star
within the error circle (Figure~3) is the only one with broad
emission lines and other spectral features that are
characteristic of a reddened Be star.  Be stars are often hard
X-ray sources because of emission from an unseen compact
companion such as a neutron star.  No other star in the error
box has optical spectroscopic properties resembling those of
known hard X-ray sources.  On this basis we identify the Be
star with SAX~J0635+0533.  This star is listed in the USNO
A1.0 astrometric catalog of the Palomar Observatory Sky Survey
(Monet et al. 1996), at position (J2000) RA = $06^{\rm
h}35^{\rm m}18^{\rm s}\!.29$, Dec =
$+05^{\circ}33^{\prime}6^{\prime\prime}\!.3$.  The B1950
coordinates are RA = $06^{\rm h}32^{\rm m}38^{\rm s}\!.11$,
Dec = $+05^{\circ}35^{\prime}34^{\prime\prime}\!.3$.  The same
catalog gives approximate magnitudes of $B = 12.7$ and $R =
12.7$. There is no 20~cm radio source present at this position
in the NVSS catalog (Condon et al. 1998).

Additional optical spectra were obtained with the Goldcam
spectrograph on the KPNO 2.1~m telescope on 1998 January 28,
29, and 30.  On the first two nights the spectra covered the
wavelength range 3867--7516~\AA\ with a resolution of
5--6~\AA, while on the third night the spectra covered the
range 5548--8613~\AA\ with a resolution of 4~\AA. Three
sequences of 300~s exposures were taken on each night (at the
beginning middle, and end of the night), yielding a total of
39 spectra. Lastly, we obtained one more spectrum on the
Hiltner telescope on 1998 April 7.

All of the spectra obtained are consistent with each other in
continuum flux and emission-line properties.  The mean of
several of the KPNO spectra is shown in Figure~4, and their
emission-line properties are listed in Table~1. The quoted
line widths have been corrected for instrumental resolution.
The strong H$\alpha$ emission line has a narrow core and very
broad wings. It can be fitted well by the sum of a broad and a
narrow Gaussian.  The narrow component has an equivalent width
(EW) of 27~\AA\ and a full-width at half maximum (FWHM) of
330~km~s$^{-1}$, while the broad component has EW = 8~\AA\ and
FWHM = 1300~km~s$^{-1}$.  The wings of H$\alpha$ extend to a
full width at zero intensity (FWZI) of
$2100\pm200$~km~s$^{-1}$.

A 300~s exposure on SAX~J0635+0533 was obtained by Goeff Marcy
using the HIRES instrument (Vogt et al. 1994) on the Keck~I
telescope, on 1998 January 26. The resolution is
2~km~s$^{-1}$.  Figure~5 shows the profiles of H$\beta$ and
He~I~$\lambda$5876, after normalizing them by the local
continuum level and smoothing with a 14~km~s$^{-1}$ running
boxcar filter.  The emission lines are double peaked, with
peak separation of $\approx 200$~km~s$^{-1}$ and FWZI $\approx
500$~km~s$^{-1}$.  Thus it appears that the narrow component
identified in the moderate-resolution spectra as having FWHM
$\approx 330$~km~s$^{-1}$ is actually resolved into a
disk-like profile, similar to what is seen in the Be star
X-ray transient A0535+26 by Clark et al. (1998), and
attributed by them to a circumstellar disk around the Be star.

The individual KPNO spectra were used to search for radial
velocity variations of the H$\alpha$ line over the three
nights, but no significant variations were found. The scatter
in the measured velocities appears to be random, and defines
an upper limit to the real radial velocity variations on this
time scale of 30~km~s$^{-1}$.

\section{Mean Spectrum, Reddening, and Distance} 

The {\it observed} Johnson magnitudes (Johnson \& Morgan 1951)
measured from the mean spectrum of Figure~4 are $B=13.81$,
$V=12.83$, $R=11.98$.  Absorption in many diffuse interstellar
bands (DIBs) is apparent in Figure~4, and the equivalent
widths of these can be used to estimate reddening and
distance.   We measured four of the best-studied interstellar
absorption features; their EWs are listed in Table~1.  In
addition to the EWs, the depth of the $\lambda 4430$ feature
was also measured to be 13\% below the continuum (this
quantity is a reddening indicator in addition to the EW).
There are several calibrations of the DIBs in the literature. 
The most extensive work is that of Herbig (1975) who studied
the correlation of the EWs of many DIBs with reddening. 
Wampler (1963, 1966) noted that the calibration of the
$\lambda 4430$ feature as a reddening indicator depends on
Galactic longitude.  More recently, T\"ug \& Schmidt-Kaler
(1981) also investigated this issue and came up with their own
calibration for the depth of the $\lambda 4430$ feature. We
estimated the extinction $E(B-V)$ to SAX~J0635+0533 using each
of these calibrations, and adopted a mean value of $E(B-V)=1.2
\pm 0.2$, where the uncertainty corresponds to the dispersion
among different calibrations.  Assuming that $A_{\rm V} =
3.2\, E(B-V)$, we find $A_{\rm V} = 3.8 \pm 0.6$.

The Mon~OB2 association is a cluster of massive stars at a
distance of 1620~pc (Garmany \& Stencel 1992), and within
$1^{\circ}$ of SAX~J0635+0533.  The H~I column densities to 11
stars in this cluster have been measured with IUE using Lyman
$\alpha$ absorption and are in the range $1.4-3.8 \times
10^{21} \rm \, cm^{-2}$ (Diplas \& Savage 1984).  The
estimated $E(B-V)$ to SAX~J0635+0533 corresponds to $N_{\rm H}
= 6.0 \times 10^{21}$~cm$^{-2}$ using standard conversions
(Savage \& Mathis 1979).  Thus we can be confident that
SAX~J0635+0533 is no closer than the 1620~pc distance to Mon
OB2.  The mean column density to the 11 stars is $1.93 \times
10^{21} \rm \, cm^{-2}$.  Assuming that $A_{\rm V} = 1$ for
$N(H~I) = 1.6 \times 10^{21} \rm \, cm^{-2}$, this corresponds
to $A_{\rm V} = 0.8$~mag/kpc.  If we assume that 0.8 mag/kpc
is a typical value along the line of sight, the distance to
SAX~J0635+533 is probably not further than 5~kpc.

The total Galactic 21~cm column density in this direction is
$7.0 \times 10^{21}$~cm$^{-2}$ (Stark et al.\ 1992), which is
larger than the column density estimated for SAX~J0635+0533
from interstellar absorption.  Thus, we regard the estimated
extinction from the DIBs as compatible with the 21~cm H~I
column, while the larger X-ray measured value, $N_{\rm H} =
(2.0 \pm 0.3) \times 10^{22}$~cm$^{-2}$, implies that there
must be some circumstellar gas that is responsible for the
extra X-ray absorption.  We note that the emission-line ratio
of H$\alpha$/H$\beta$ can be used with recombination theory to
derive an upper limit to $E(B-V)$ of 1.4, and that this upper
limit is also well below the X-ray measured column.

After correction for reddening, the Johnson magnitudes
estimated from the spectrum become $B=8.74$, $V=8.99$, and
$R=9.14$.   Such colors are characteristic of an early B type
star.  For a distance range of 2.5--5~kpc, the absolute visual
magnitude would be in the range $-3.1$ to $-4.5$, and the
stellar classification in the range  B2V to B1III.

\section{Discussion}

Assuming 200 Be/X-ray binaries within 5~kpc (Rappaport \& van
den Heuvel 1982) distributed uniformly in Galactic longitude
and  in a latitude distribution with an rms width of
$2.2^{\circ}$, we estimate there is a 4\% chance of finding a
spurious overlap of a Be/X-ray binary in an error box the size
of that for 2EG~J0635+0521.  This probability is reasonably
low.  In addition, both Be/X-ray binaries and the low-latitude
EGRET sources are preferentially found near OB associations
(Kaaret \& Cottam 1996). This may suggest that some of the
unidentified EGRET sources are, in fact, associated with
Be/X-ray binaries.

While identification of SAX~J0635+0533 and its optical
counterpart with 2EG~J0635+0521 must remain tentative awaiting
an improved gamma-ray position or detection of periodicity in
both x-rays and gamma-rays, the positional coincidence and the
hard spectrum of the x-ray source are suggestive and here we
entertain some speculations based on the possible
identification.   Assuming the association of the x-ray source
SAX~J0635+0533 with the optical counterpart discussed above
and with 2EG~J0635+0521, allows us to find the luminosity of
the source based on the optical distance estimate.  The 2--10
keV X-ray luminosity would be $9 - 35 \times 10^{33} \rm \,
ergs \, s^{-1}$ and the gamma-ray luminosity for 0.1--5~GeV
would be $1.3 - 5 \times 10^{35} \rm \, ergs \, s^{-1}$, both
assuming isotropic emission.   The x-ray to gamma-ray flux
ratio is $0.07$.  If the identification is correct, then the
source would be a gamma-ray emitting x-ray binary.

It is possible that the binary contains a neutron star and
that both the x-ray and gamma-ray emission are magnetospheric
pulsar emission.  If the binary is wide, as is tentatively
suggested by our non-detection of radial velocity variations,
then the pulsar emission may be be relatively uninfluenced by
the presence of the companion star.  There are several pieces
of evidence in favor of pulsar emission.  The ratio of x-ray
to gamma-ray flux is consistent with those of ``Vela-like''
pulsars.   The power-law x-ray spectrum with photon index near
1.5 is similar to that of most isolated gamma-ray pulsars
above 2~keV (Wang et al. 1998).  The gamma-ray flux is
constant over 14 different observations (McLaughlin et al.
1996), as would be expected for a pulsar.  And finally, as
noted above, 2EG~J0635+0521 was identified by Merck et al.
(1996) as one of only eight unidentified Galactic gamma-ray
sources with energy spectra sufficiently hard to be pulsar
candidates.

Alternatively, either the x-ray emission, gamma-ray emission,
or both may arise from interaction of the energetic particle
wind from the pulsar with the wind or the radiation from the
companion.  Such mechanisms have been suggested for
Be/millisecond radio pulsar binary PSR~J1259-63 and the
variable radio Be/X-ray binary system LSI~61$^{\circ}$~303. 
In PSR~J1259-63, emission up to 200~keV  is thought to arise
from a shock interaction of the energetic particle wind from a
pulsar with the wind from the Be-star (Grove et al. 1995). 
The x-ray photon index, 1.5--1.9 (Kaspi et al. 1995), is
similar to that we find for SAX~J0635+0533.  However,
PSR~J1259-63 was not detected by EGRET (Tavani et al. 1996)
and the x-ray emission is highly variable.  It is possible
that the shock acceleration mechanism could operate to higher
energies if the shock region is less compact than that of
PSR~J1259-63 at periastron (Tavani \& Arons 1997).

The gamma-ray source 2CG~135+01 has long had a tentative
association with the unusually variable radio source and
Be/X-ray binary LSI~61$^{\circ}$~303 (Gregory \& Taylor 1978;
for recent work see Strickman et al. 1998).  It has been
suggested (Maraschi \& Treves 1981) that the source contains a
young pulsar and that the gamma-rays arise from
inverse-Compton scattering of the optical photons from the
B-star in the relativistic wind from the pulsar at the
boundary between the pulsar wind and the stellar wind. 
However, 2CG~135+01 is a variable gamma-ray source (Tavani et
al. 1998).  If 2EG~J0635+0521 is similar to 2CG~135+01, then
the constancy of its gamma-ray flux (McLaughlin et al. 1996)
would suggest a low eccentricity orbit.

Finally, it may be that the emission is powered not by
rotation of a neutron star, but rather by accretion.  The
recently reported detection of GeV pulsations from Cen~X-3
with a frequency equal to that detected contemporaneously in
x-rays (Vestrand, Sreekumar, \& Mori 1997), albeit not with an
outstandingly high significance, suggests the possibility that
2EG~J0635+0521 could be an accretion powered system.  The
mechanism for gamma-ray production could be shock acceleration
or a magnetospheric process.  If SAX~J0635+0533 is similar to
Cen~X-3, the x-ray emission would be from accretion and should
be pulsed.  However, the lower bound we obtained for the
exponential cutoff energy of the x-ray spectrum is
significantly higher than the cutoffs typically found in
high-mass x-ray binaries (White, Swank, \& Holt 1983) and
would be more suggestive of either a low magnetic field
neutron star or a black hole.

Identification of SAX~J0635+0533 with the gamma-ray source
2EG~J0635+0521 would argue against interpreting the gamma-ray
emission as due to cosmic-ray production via shock
acceleration in the supernova remnant (Esposito et al. 1996;
Jaffe et al. 1997).  We note that we detect no diffuse hard
x-ray emission in the BeppoSAX data.

Possible ways to strength the identification of the
x-ray/optical source and the gamma-ray source would be to
detect a periodicity in the x-ray, optical, or radio which
could then be detected in gamma-rays or to extend the x-ray
spectrum and join it to the gamma-ray spectrum.  Finally, if
the gamma-ray emission extends to higher energies, an improved
gamma-ray position may be obtained from a ground-based TeV
telescope such as the Whipple.

\acknowledgments

We thank Fabrizio Fiore and the Sax Data Center for assistance
with the SAX data.  We thank Arlin Crotts for obtaining
identification spectra at MDM Observatory, and Geoff Marcy for
obtaining the Keck HIRES spectrum.  P.~K. acknowledges support
from NASA grant NAG5-7389.  M.~E. acknowledges support from
Hubble fellowship grant HF-01068.01-94A from Space Telescope
Science Institute, which is operated for NASA by the
Association of Universities for Research in Astronomy, Inc.,
under contract NAS~5-26255.


\newpage

\begin{deluxetable}{lcccc}
\tablecaption{Lines Measured from the Spectrum}
\tablewidth{0pt}
\tablehead{\colhead{Feature} & 
           \colhead{EW} & 
           \multicolumn{2}{c}{Flux $(\rm erg \, cm^{-2} \, s^{-1})$} &
           \colhead{FWHM}\nl
           &   
           \colhead{(\AA)} &  
           \colhead{Observed} & 
           \colhead{Dereddened\tablenotemark{a}} &
           \colhead{$(\rm km \, s^{-1})$} }

\startdata

\multicolumn{4}{c}{Emission Features} & \nl
H$\beta$                      & \phantom{0}3.22 & $6.69\times 10^{-14}$ & $3.91\times 10^{-12}$ & 400 \nl
He~{\sc i} $\lambda$5876      & \phantom{0}1.31 & $4.67\times 10^{-14}$ & $1.13\times 10^{-12}$ & 600 \nl
H$\alpha$                     & 30.3\phantom{0} & $8.89\times 10^{-13}$ & $1.39\times 10^{-11}$ & 390 \nl
He~{\sc i} $\lambda$6678      & \phantom{0}0.40 & $1.18\times 10^{-14}$ & $1.74\times 10^{-13}$ & 450 \nl
He~{\sc i} $\lambda$7064      & \phantom{0}0.84 & $2.57\times 10^{-14}$ & $3.17\times 10^{-13}$ & 470 \nl
O~{\sc i} $\lambda$8446, Pa18 & \phantom{0}3.35 & $8.94\times 10^{-14}$ & $6.39\times 10^{-13}$ & 490 \nl
Pa17                          & \phantom{0}2.10 & $5.51\times 10^{-14}$ & $3.89\times 10^{-13}$ & 570 \nl
Pa16                          & \phantom{0}2.72 & $7.08\times 10^{-14}$ & $4.88\times 10^{-13}$ & 490 \nl
Pa15                          & \phantom{0}1.73 & $4.51\times 10^{-14}$ & $3.03\times 10^{-13}$ & 470 \nl

\multicolumn{4}{c}{Absorption Features} & \nl
I.S. $\lambda$4430                     & $2.9\pm 0.1$  & & & \nl
I.S. $\lambda \lambda$5778,5780,5797   & $2.5\pm 0.2$  & & & \nl
Na~{\sc i} $\lambda \lambda$5890,5896  & $2.1\pm 0.2$  & & & \nl
I.S. $\lambda$6283                     & $2.0\pm 0.1$  & & & \nl

\tablenotetext{a}{Assuming $E(B-V) = 1.2$.}
\enddata
\end{deluxetable}


\begin{references}

\reference{} Bignami, G.F. \& Hermsen, W. 1983, \araa, 21, 67

\reference{} Clark, J. S., et al. 1998, \mnras, 294, 165

\reference{} Condon, J. J., Cotton, W. D., Greisen, E. W.,
Yin, Q. F., Perley, R. A., Taylor, G. B., \& Broderick, J. J.
1998, \aj, 115, 1693

\reference{} Diplas, A. \& Savage, B.D. 1984, \apjs, 93, 211

\reference{} Esposito, J.A., Hunter, S.D., Kanbach, G., \&
Sreekumar, P. 1996, \apj, 461, 820

\reference{} Garamany, C.D. \& Stencel, R.E. 1992, \aaps, 94,
214

\reference{} Gregory, P.C. \& Taylor, A.R. 1978, Nature, 272,
704

\reference{} Grove, J.E., Tavani, M., Purcell, W.R., Johnson,
W.N., Kurfess, J.D., Strickman, M.S., \& Arons, J. 1995,
\apjl, 477, L113

\reference{} Halpern, J.P.  \& Holt, S.S. 1992, Nature, 357,
222

\reference{} Herbig, G. 1975, \apj, 196, 129

\reference{} Jaffe, T.R., Bhattacharya, D., Dixon, D.D., \&
Zych, A.D. 1997, \apjl, 484, L129

\reference{} Johnson, H.L. \& Morgan, W.W. 1951, \apj, 113, 522


\reference{} Kaaret, P. \& Cottam, J. 1996, \apj, 462, L35

\reference{} Kaspi, V.M., Tavani, M., Nagase, F., Hirayama,
M., Hoshino, M., Aoki, T., Kawai, N., \& Arons, J. 1995, \apj,
453, 424

\reference{} Leahy, D.A., Naranan, S., \& Singh, K.P. 1986,
MNRAS, 220, 501

\reference{} Maraschi, L. \& Treves, A. 1981, MNRAS, 194, 1P

\reference{} McLaughlin, M.A., Mattox, J.R., Cordes, J.M., \&
Thompson, D.J. 1996, \apj, 473, 763

\reference{} Merck, M. et al.\ 1996, \aaps, 120, 465

\reference{} Monet, D. et al. 1996, USNO-SA1.0 (Washington,
DC: US Naval Observatory)

\reference{} Ramanamurthy, P.V., et al. 1995, \apj, 447, L109

\reference{} Rappaport, S.  \& van den Heuvel, E.P.J. 1982, in
Proc. IAU Symposium 98, ``Be Stars'', (eds Jaschek, M. \&
Groth, H.-G.) p. 327

\reference{} Savage, B.D. \& Mathis, J.S. 1979, \araa, 17, 73

\reference{} Stark, A.A., Gammie, C.F., Wilson, R.W., Bally,
J., Linke, R.A., Heiles, C., \& Hurwitz, M. 1992, \apjs, 79,
77

\reference{} Strickman, M.S. et al. 1998, \apj, 497, 419

\reference{} Swanenburg, B.N., et al. 1981, \apj, 243, L69

\reference{} Tavani, M. \& Arons, J. 1997, \apj, 477, 439

\reference{} Tavani, M., Kniffen, D., Mattox, J.R., Paredes,
J.M., Foster, R. 1998, \apj, 497, L89

\reference{} Thompson, D.J., et al. 1994, \apj, 436, 229

\reference{} Thompson, D.J., et al. 1995, \apjs, 101, 259

\reference{} Torii, K. et al. 1997, \apj, 489, L145

\reference{} Torii, K. et al. 1998, \apj, 494, L207

\reference{} T\"ug, H., \& Schmidt-Kaler, Th. 1981, \aap, 94, 16

\reference{} Vestrand, W.T., Sreekumar, P. \& Mori, M. 1997,
\apj, 483, L49

\reference{} Vogt, S.S. et al.\ 1994, in Proc. SPIE Vol. 2198,
``Instrumentation in Astronomy VIII'', (eds Crawford, D.L. \& 
Craine, E.R.), p. 362

\reference{} Wade, R. A., \& Oke, J. B. 1977, \apj, 215, 568

\reference{} Wampler, J. E. 1963, \apj, 137, 1071

\reference{} Wampler, J. E. 1966, \apj, 144, 921

\reference{} Wang, F.Y.-H., Ruderman, M., Halpern, J.P., \&
Zhu, T. 1998, \apj, 498, 373

\end{references}
\end{document}